\documentclass[a4paper,fleqn]{cas-dc}

\usepackage[numbers]{natbib}
\usepackage[percent]{overpic}
\usepackage{siunitx}
\usepackage[linesnumbered,ruled,vlined]{algorithm2e}

\def\tsc#1{\csdef{#1}{\textsc{\lowercase{#1}}\xspace}}
\tsc{WGM}
\tsc{QE}
\tsc{EP}
\tsc{PMS}
\tsc{BEC}
\tsc{DE}

\newcommand{\change}[1]{{#1}}

\begin{document}
\let\WriteBookmarks\relax
\def\floatpagepagefraction{1}
\def\textpagefraction{.001}
\shorttitle{Naive Bayes Orphan Adoption}
\shortauthors{T Olsson et~al.}

\title [mode = title]{To Automatically Map Source Code Entities to Architectural Modules with Naive Bayes} 




\author[]{Tobias Olsson}[type=editor,
                        auid=000,bioid=1,
                        orcid=0000-0001-7511-2910]
\cormark[1]
\fnmark[1]
\ead{tobias.olsson@lnu.se}


\address[]{Department of Computer Science, Linnaeus University, Kalmar/Växjö, Sweden}

\author[]{Morgan Ericsson,}[]

\author[]{Anna Wingkvist}[]





\begin{abstract}
\noindent \textbf{Background:} The process of mapping a source code entity onto an architectural module is to a large degree a manual task. Automating this process could increase the use of static architecture conformance checking methods, such as reflexion modeling, in industry. Current techniques rely on user parameterization and a highly cohesive design. A machine learning approach would potentially require less parameters and better use of the available information to aid in automatic mapping. \textbf{Aim:} We investigate how a classifier can be trained to map from source code to architecture modules automatically. This classifier is trained with semantic and syntactic dependency information extracted from the source code and from architecture descriptions. The classifier is implemented using multinomial naive Bayes and evaluated. \textbf{Method:} We perform experiments and compare the classifier with three state-of-the-art mapping functions in eight open-source Java systems with known ground-truth-mappings. \textbf{Results:} We find that the classifier outperforms the state-of-the-art in all cases and that it provides a useful baseline for further research in the area of semi-automatic incremental clustering. \textbf{Conclusions:} We conclude that machine learning is a useful approach that performs better and with less need for parameterization compared to other approaches. Future work includes investigating problematic mappings and a more diverse set of subject systems.
\end{abstract}




\maketitle

\section{Introduction}


A software system's architecture captures the fundamental, high-level design decisions that affect the system's qualities. System qualities are often in conflict, and the architecture can help find a balance and serve as a guide for many of the decisions made during system development and evolution\change{~\cite{SAInPractice}}. However, developers face many challenges, including poorly communicated or non-existent architectural documentation, the pressure to deliver features quickly, and infeasible architectural decisions. Given enough time, it is likely that an implementation will no longer conform to its architecture if no explicit effort is spent on checking their mutual conformance\change{~\cite{SAInPractice, ControllingSAESurvey}}. \change{Architecture conformance can be checked statically at various points in the development process. A popular approach is Reflexion modeling~\cite{ControllingSAESurvey,murphy1995software}, where dependencies in the architecture are checked against those found in the implementation. The basic premise of Reflexion modeling is that there exists a designed, intended architecture with modules and their respective dependencies and that there exists a mapping from the source code to these modules.}

Yet, Static Architecture Compliance Checking (SACC) is not a common industry practice~\citep{Ali2017ArchitectureRequirements}. Consequently, long-lived systems often experience architectural erosion or drift, and in turn, they do not obtain their desired internal and external qualities~\citep{ControllingSAESurvey}. Studies show that a major reason why SACC has not reached widespread adoption is not a lack of methods or tools~\cite{JITTAC, KnodelThesis, SAVELifeExperiment}, but rather that it requires a mapping from the source code entities to the modules of the intended architecture. Generally, such a mapping does not exist, and if it does, it is not actively maintained. To create or even validate a mapping manually requires too much effort~\cite{Ali2017ArchitectureRequirements, Bittencourt2010ImprovingTechniques, Christl2007AutomatedMethod}. \change{So, automatic support for creating such a mapping is a promising approach to increase the usefulness of SACC in practice.}

The state-of-the-art in automated mapping relies on either syntactic or semantic information \change{to compute the attraction of an unmapped source code entity to the modules and then, if the attraction is found valid, an automatic mapping can be performed}. \change{\citeauthor{Christl2005EquippingClustering}}, for example, extract a dependency graph from the system and rely on the number of dependencies between entities to determine the \change{attractions~\cite{Christl2005EquippingClustering}}. \change{\citeauthor{Bittencourt2010ImprovingTechniques}} \change{instead rely} on information retrieval techniques, such as latent semantic indexing, to \change{compute the attraction by using (document) similarity~\cite{Bittencourt2010ImprovingTechniques}}. 

We rely on an approach that combines multiple sources of information from the source code, such as the dependency graph and identifier names, and use these \change{as features} to train a \change{machine learning} classifier. We first introduced the idea of using a Naive Bayes classifier to map from implementation to intended architecture in~\cite{OlssonNaiveBayes}. \change{We build on and extend our previous study by adding the two information retrieval techniques described by \citeauthor{Bittencourt2010ImprovingTechniques} to our comparison and two more subject systems with ground truths created by system experts~\cite{Bittencourt2010ImprovingTechniques}. We present a complete evaluation of the state-of-the-art and evaluate how syntactic and semantic information can be combined for the three functions when it is relevant. In addition, we improved the performance of the original work by~\citeauthor{Christl2007AutomatedMethod} by using appropriate weights for different types of dependencies in the dependency graph~\cite{Christl2007AutomatedMethod}}.

\change{The goals of our study is to evaluate: 
\begin{itemize}
\item How does the combination of syntactic and semantic information affect the performance of text-based attraction functions in general? 
\item How well does our Naive Bayes classifier perform compared to the state-of-the-art?
\end{itemize}
}



Our main contribution is to show how machine learning can be a viable approach to automated mapping from implementation to intended architecture. Our approach outperforms the current state-of-the-art and requires a smaller initial set and much fewer parameters. We also show how syntactic information in the form of dependencies can be combined with semantic information to improve performance, both for our classifier and the techniques by \citet{Bittencourt2010ImprovingTechniques}. \change{Finally, we contribute with an open-source implementation~\cite{Olsson2021} and an in-depth analysis of the initial set and parameterization of the attraction functions.}

The remainder of the article is organized as follows. We begin by introducing the state-of-the-art in automated mapping, including our approach. We continue with \change{a description of} how the experiment is conducted as well as how data is collected and analyzed. We also discuss how we implemented \change{the} mapping algorithms. We then present our results that visually and quantitatively show the difference in the performance of different \change{attraction functions} and our analysis \change{of these}. Finally, we present related work, our conclusions, and future work.

\section{Background}\label{sec:background}

\change{In this work, we consider the structural architecture, i.e., how the source code is organized into cohesive modules that can be reasoned about regarding qualities such as reuse, performance, maintainability etc.} 

We need a mapping from the source code to the architecture to reason about how well an implementation conforms to the intended architecture using, for example, Reflexion modeling~\cite{murphy1995software}. \change{The mapping approach is semi-automatic}, where the mapping is created using an initial set of mapped source code entities and an intended modular architecture.

The source code model consists of \emph{Entities} (E) and \emph{Dependencies} (ED). The entities are constructs, such as classes, defined in a programming language, and the ED result from, e.g., method calls and inheritance. The architecture model consists of \emph{Modules} (M) and (allowed) \emph{Module Dependencies} (MD) between these. The modules represent major parts of the architecture of a system, e.g., the GUI and the logic. The directed MDs indicate how these modules are allowed to interact and depend on each other. 

\begin{figure}[]
    \centering
    \includegraphics[width=0.98\columnwidth]{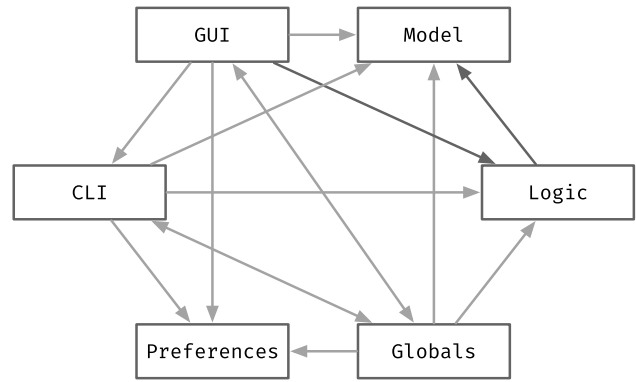}
    \caption{The intended architecture of JabRef version 3.7. The boxes represent modules, and the arrows represent allowed directed dependencies between these modules.\label{fig:jabref_intended}}
\end{figure}

Figure~\ref{fig:jabref_intended} depicts the intended architecture of one of our subject systems, JabRef version 3.7~\cite{SAEroConJabRef}. There are six modules and thirteen dependencies between them. For example, \change{\textsf{GUI}} depends on \change{\textsf{Logic}} and \change{\textsf{Globals}}, and \change{\textsf{Logic}} depends on \change{\textsf{Model}} (darker gray arrows). The required functionality of these modules is implemented by several entities, in this case Java classes.

\begin{figure}[]
    \centering
    \includegraphics[width=0.95\columnwidth]{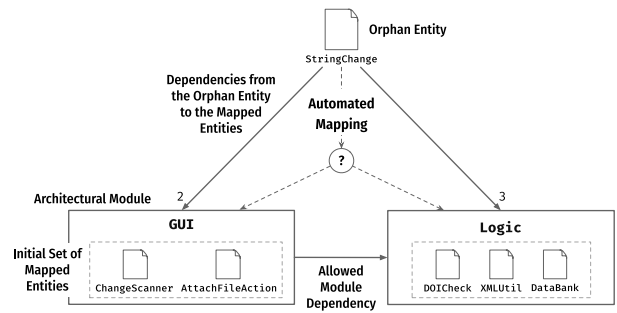}
    \caption{An example of how entities are mapped to modules. In this case, two classes are mapped to GUI, and three classes are mapped to Logic. These form the initial set. A sixth class, StringChange, is about to be mapped. \change{It has two dependencies to entities in the GUI module, and three dependencies to the entities in the Logic module.}\label{fig:mapping_ex}}
\end{figure}

The objective of an automated mapping algorithm is to \emph{map} each \emph{Entity} to the correct \emph{Module}. Figure~\ref{fig:mapping_ex} depicts a subset of the JabRef modules and entities. In this example, the mapping algorithm's objective is to decide which of the two modules, \change{\textsf{GUI}} and \change{\textsf{Logic}}, the entity \change{\textsf{StringChange}} should be mapped to. Once such a mapping exists for all entities, we can compare the ED of the implementation to the MD allowed by the architecture to determine whether they are convergent, absent, or divergent~\cite{murphy1995software}. For example, if we map \change{\textsf{StringChange}} to \change{\textsf{Logic}, the two dependencies from \textsf{StringChange} to \textsf{GUI} would be considered divergent (cf.~\ Figure~\ref{fig:mapping_ex}).}

\subsection{Semi-automatic Incremental Clustering}

We rely on \textit{orphan adoption}~\cite{tzerpos1997orphan}, an incremental clustering technique, to map entities to modules automatically. An unmapped entity is considered an orphan that should be adopted by one of the modules. For example, in Figure~\ref{fig:mapping_ex}, the entity \change{\textsf{StringChange}} is an orphan that should be \change{mapped to} either \change{\textsf{GUI}} or \change{\textsf{Logic}}.


\citet{tzerpos1997orphan} identify four criteria that can guide the mapping:

\begin{enumerate}
 \item \textit{Naming}: Naming standards can reveal what module is suitable.
 \item \textit{Structure}: Dependencies between an orphan and already mapped entities can be used as a mapping criterion.
 \item \textit{Style}: Modules are often created using different design principles, e.g., high cohesion or not. Classifying the orphan based on style can give hints on how to use, for example, the structure criteria.
 \item \textit{Semantics}: The source code can be analyzed to determine its purpose and how similar it is to the modules' purpose.
\end{enumerate}


\change{\citet{Christl2007AutomatedMethod} introduced the Human Guided clustering Method (HuGMe), an approach to semi-automatic mapping of source code entities to modules of the intended architecture based on orphan adoption~\cite{tzerpos1997orphan}. It is an iterative approach that, at its core, uses an attraction function to compute an attraction between an orphan and a module. It contains the following steps. Note that specific attraction functions can alter or skip certain steps as suitable.} 

\begin{enumerate}[1.]
    \item Construct an initial set of mapped entities, cf.\ Figure~\ref{fig:mapping_ex}. This is a manual process, where one or several experts create a mapping from a subset of the entities to the intended modules.\label{step:initial}
    \item Find all source code entities that are candidates for automatic mapping. These entities become the set of orphans \change{for the current iteration}. \label{step:filter}
    \item For each orphan and module, compute an attraction value \change{using an attraction function}.\label{step:attraction}
    \item If the attraction between an orphan and a single module \change{meets some criterion}, the orphan is mapped to that module.\label{step:mapping}
    \item If an orphan cannot be automatically mapped, it is considered for manual mapping, where the attraction values serve as a guideline.
    \item As long as any orphan was mapped, manually or automatically, repeat from Step~\ref{step:filter}.
    \item Manually map any remaining entities.
\end{enumerate}

The HuGMe approach is used and studied \change{with a focus on different attraction functions} by, for example, \citet{Bittencourt2010ImprovingTechniques} and \citet{OlssonNaiveBayes}.

\subsection{Initial Set and Orphan Set Construction}

\change{We have previously shown that an intelligently composed and sufficiently large initial set can significantly increase the performance of an attraction function~\cite{OlssonNaiveBayes, OlssonTowardsImprovedMapping}. For example, for small initial sets ($<10\%$), we found an increase in median automatic mapping performance from 23.83\% to 44.69\% when choosing a well-composed initial set~\cite{OlssonTowardsImprovedMapping}. This shows that it is important to consider how the} initial set is constructed (Step~\ref{step:initial} \change{of HuGMe}). A larger initial set provides more data for, e.g., a machine-learning algorithm to train on. However, it also requires a significant manual effort, so these larger initial sets are generally only available in evolutionary contexts, where few new entities are added to an existing system with a known mapping.

For small initial sets, the composition is critical: \change{the initial set needs to represent} the larger system. We previously investigated whether source code metrics could guide the composition of a small initial set and found that \change{dependency-based metrics showed some promise for one of the original HuGMe attraction functions}~\cite{OlssonTowardsImprovedMapping}. However, the results could not be generalized to a larger set of systems, so \change{using} heuristics to guide the creation is still an open issue.


\change{Orphans are the unmapped entities considered for automatic mapping in the current iteration of the process.}
A simple way to construct the set of orphans is to consider all unmapped entities as orphans. However,~\citeauthor{Christl2007AutomatedMethod} suggest additional filtering (Step~\ref{step:filter} \change{of HuGMe}) \change{for the attraction functions they present} based on the ratio between an entity's dependencies and the number of dependencies to entities already mapped. If this value exceeds a threshold $\omega$, the entity is considered an orphan. The intention is to avoid mapping orphans that are likely to be affected by other orphans' subsequent mapping. The end-user determines $\omega$, but the general advice is to keep it low if the initial set is small. 


\subsection{Attraction Functions, Training, and Mapping}

How attraction values are computed (Step~\ref{step:attraction} \change{of HuGMe}) and how they are used to determine whether an orphan should be mapped or not (Step~\ref{step:mapping} \change{of HuGMe}) are critical to how well an automated mapping method performs. \change{We outline the five previously studied attraction functions, including our Naive Bayes-based function.}

\citeauthor{Christl2007AutomatedMethod} suggest and evaluate two functions to compute attractions (Step~\ref{step:attraction} \change{of HuGMe}): \textit{CountAttract} and \textit{MQAttract}. Both rely on the assumption that modules are designed to have high cohesion, i.e., more internal than external dependencies. CountAttract strives to maximize module cohesion. It computes attraction between an orphan and a module by first determining the orphan's total number of dependencies and then subtracting the number of dependencies between the orphan and other modules. The dependencies between the orphan and other modules are divided into two categories; convergent and divergent dependencies. A weighting factor $\phi$ reduces the convergent dependency count. $\phi$ is defined in the range $[0,1]$; a value close to 0 means that divergent dependencies will be the only dependencies that penalize the attraction calculation, e.g., we trust that there is conformance to the architecture. If $\phi$ is closer to 1, we put less trust in this conformance. 

\change{Considering the extreme $\phi$ values (0 and 1)}, assume there are two modules in a system: \change{\textsf{GUI} ($G$)} and \change{\textsf{Logic} ($L$)}, and dependencies are allowed from $G$ to $L$, see Figure~\ref{fig:mapping_ex}. An orphan \change{\textsf{StringChange} ($o$)} has two dependencies to module $G$ and three dependencies to module $L$. If we do not trust that there is conformance to the architecture, i.e., set $\phi = 1$, we get the following attractions: $o_G = 5 - 3*\phi = 2$, $o_L = 5 - 2*1 = 3$. If we instead trust that there is conformance to the architecture and set $\phi = 0$, we get the following attractions $o_G = 5 - 3*\phi = 5$, $o_L = 5 - 2*1 = 3$.

Notice that the three dependencies for $o_G$ are convergent in the latter case and do not reduce the attraction as $\phi = 0$. In the former case, where overall cohesion is prioritized, the attractions suggest that $o$ should be mapped to $L$. In the latter case, where conformance to the architecture is prioritized, the attractions suggest that $o$ should be mapped to $G$. This parameterization of CountAttract is dependent on expert knowledge of the system implementation and architecture. MQAttract determines the attraction values by computing the overall modularization quality when an orphan is sequentially assigned to a module. This attraction function is based on TurboMQ, \change{a fitness function used in software clustering}~\cite{TurboMQ}. \change{We do not consider MQAttract further since CountAttract has been shown to outperform MQAttract in both the original studies and subsequent studies by \citeauthor{Bittencourt2010ImprovingTechniques}~\cite{Bittencourt2010ImprovingTechniques,Christl2005EquippingClustering,Christl2007AutomatedMethod}}.


In Step~\ref{step:mapping} \change{of HuGMe}, \citeauthor{Christl2007AutomatedMethod} propose two conditions for when an orphan can be automatically mapped to a module: 1.\ \change{A single} module has an attraction value that is larger than the arithmetic mean of all of the orphan’s attraction values; or 2.\ \change{a single} module has an attraction value that is one standard deviation above the arithmetic mean of all \change{of the orphans'} attraction values. If either \change{are true}, the orphan is mapped to the corresponding module.

\citeauthor{Bittencourt2010ImprovingTechniques} rely on information retrieval techniques to compute attraction values~\cite{Bittencourt2010ImprovingTechniques}. The source code entities are considered documents, and their filenames and the names of identifiers in them as the document vocabulary. The modules form documents with vocabularies that consist of the module name and the vocabularies of all entities mapped to the module. These module documents initially contain the vocabulary of the entities in the initial set. The vocabulary can be extended as additional entities are mapped to a module. \change{For example, the orphan \textsf{StringChange} and the mapped entity \textsf{ChangeScanner} both contain the word \textsf{Change} that indicate some semantic relation (cf. Figure~\ref{fig:mapping_ex}).} \citeauthor{Bittencourt2010ImprovingTechniques} define two attraction functions based on common information retrieval techniques: \textit{IRAttract} and \textit{LSIAttract}. IRAttract uses the cosine similarity between an orphan document and a module document as the attraction value, and LSIAttract uses Latent Semantic Indexing (LSI)~\cite{deerwester1988improving} to find module documents that are semantically similar to an orphan document.

Our approach, \textit{NBAttract}~\cite{OlssonNaiveBayes}, is similar to that of \citeauthor{Bittencourt2010ImprovingTechniques}, but relies on text classification rather than on similarity measures. We define our attraction function based on Naive Bayes classification, a probabilistic approach that uses Bayes theorem and assumes that individual term probabilities are independent. We construct the module and entity documents like \citeauthor{Bittencourt2010ImprovingTechniques}, but also consider an entity's abstract dependencies to form the text using \emph{Concrete Dependency Abstraction (CDA)}. The classification uses the terms of the documents in the training data, which is the initial set, to build a probabilistic model of the classifications. This model can then be used to assign a probability of an orphan belonging to each module. If the highest probability is considered high enough, e.g., above a predefined threshold, the orphan belongs to the module.

\change{
As neither IRAttract, LSIAttract, nor NBAttract relies on program dependencies, the filtering approach suggested by \citeauthor{Christl2007AutomatedMethod} does not apply. For these attraction functions, all unmapped entities are considered orphans in Step ~\ref{step:filter} \change{of HuGMe}.
}

\subsection{Concrete Dependency Abstraction}

CDA incorporates syntactical information of dependencies as document terms. This way, syntactic and semantic information can be combined without the need for additional algorithms, weights, or ordering of two or more algorithms~\cite{Bittencourt2010ImprovingTechniques}. CDA forms document terms based on dependencies in the source code. CDA terms are generated based on the source module's name, the type of dependency, and the target module's name. In essence, the dependency  abstracts from the implementation level (between two entities) to the architectural level (between two modules) and keeps the type of dependency. For example, if entity $e_1$ inherits from $e_2$, $e_1$ is mapped to $A$, and $e_2$ is mapped to $B$, then the term $A-Inherits-B$ is generated and added to the documents of both modules $A$ and $B$. Adding these terms will promote adding future entities to $A$ or $B$, \change{if the entities have dependencies that also gives CDA} inheritance dependencies between these two modules.

To classify an orphan, we create one document for each module that hypothetically maps the orphan to that module. The terms in each of these documents are created as if the orphan was mapped to that particular module \change{and can be combined with other textual terms such as names of identifiers etc. Each unique orphan-module document is then used to calculate the probability of the orphan belonging to a particular module, effectively calculating the similarity of the orphan to the module as if the orphan was mapped to the module.} 

Continuing the example above, assume an orphan $o$ also inherits from an entity in $B$. The abstract dependencies for $o$ as if mapped to $A$ generates the string $A-Inherits-B$ and the abstract dependencies of $o$ as if mapped to $B$ generates the string $B-Inherits-B$. In this example, mapping $o$ to $A$ is promoted as $A$ also includes the term $A-Inherits-B$ in its document.

The CDA terms can be used together with any attraction function that relies on textual terms. This is an advantage compared to \textit{CountAttract}, that only uses \change{syntactical information of }dependencies and \citeauthor{Bittencourt2010ImprovingTechniques}, that use only the semantic information \cite{Bittencourt2010ImprovingTechniques}. \change{Another advantage is} that CDA uses the actual dependencies and types of dependencies found in the implementation, i.e., if there are divergent dependencies. \change{In addition}, there is no need for individual weighting of different types of dependencies as in CountAttract. A machine learning model should be able to use these advantages when determining the mapping and perform better with less parameterization.




\section{Method}

\change{The goals of our study are to evaluate: 1.\ How does CDA affect the performance of the text-based attraction functions IRAttract, LSIAttract and NBAttract 2.\ How well does the NBAttract attraction function perform compared to CountAttract, IRAttract and LSIAttract? We extend our previous study~\cite{OlssonNaiveBayes} that only compared CountAttract and NBAttract. We also add an investigation of the effect of CDA.}

We implement four attraction functions: CountAttract, IRAttract, LSIAttract, and NBAttract. We omit MQAttract since other studies~\cite{Bittencourt2010ImprovingTechniques, Christl2005EquippingClustering, Christl2007AutomatedMethod} found that CountAttract outperforms it. 


\change{We study open-source systems, of which we know the intended architecture, i.e., we have mappings between the architecture and the respective source code entities.} We randomly generate a large number of initially mapped sets of different sizes and compositions from these systems and apply the attraction functions to map the remaining orphans. We follow the \change{HuGMe} algorithm specified in Section~\ref{sec:background} and iteratively run the attraction function as long as an orphan was mapped in the previous iteration. Once the algorithm has finished, we compare the result to the known mapping and compute precision and recall. All parameters of the attraction functions, e.g., whether CDA should be used or the $\phi$-value, are randomly generated for each run of the experiment. \change{One exception is the weights for individual dependency types used by CountAttract. These are approximated by a genetic optimization algorithm.} Note that there is no human intervention; we \change{neither} map hard cases \change{manually nor} correct the automatic mapping between iterations. As a result, the automatic mappings may contain errors and some orphans may be left unmapped.

\change{
We justify using random initial sets because an attraction function should be robust regardless of the size and the composition of the initial set. There is no established way of composing an initial set that generalizes to many systems or attraction functions. The individual attraction functions also rely on random parameters. Pairwise comparison for CDA, in the case of NBAttract, IRAttract, and LSIAttract, would be possible using the same initial sets. However, this would require a different analysis technique or a lot more data when comparing to CountAttract. Conceptually it is easier to use the same statistical tests, and the statistical power should be sufficient for a comparison given a large enough data set.
} 

\change{
The weights for dependencies used in CountAttract are decided by a genetic optimization algorithm for each system. The search space for combinations of continuous weights over 11 different dependencies creates a combinatorial explosion that is too large to be handled by randomization alone. Also, the filtering of the data would be very complicated. 
}

\subsection{Experiment Outline}

The experiment is performed according to Algorithm~\ref{algo:mapoutline}. A system \emph{s} consists of source code entities, architecture modules and allowed dependencies between these, and a ground truth mapping from entities to modules. 

\begin{algorithm}
  \DontPrintSemicolon 
  \ForEach{$s \in \textrm{Systems}$}{
    \Repeat{$\textnormal{50\,000 times}$}{
      $MI \gets \textrm{generateInitialMapping}(s)$\;
      \ForEach{$af \in Attraction Functions$}{
        setParameters($af$)\;
        $M \gets MI$\;
        \Repeat{$|M| \textnormal{ does not grow}$}{
          initialize$(af, M)$\;
          $O \gets \textrm{findOrphans}(af, M, s)$\;
          \ForEach{$o \in O$}{
            calculateAttractions$(af, M, o, s)$
          }
          \ForEach{$o \in O$}{
            $m \gets \textrm{candidateModule}(af, o, s)$\;
            \If{m}{ 
              map$(o, m)$\;
              add$(o, M)$\;
            }
          }
        }
      }
      saveData$(af, M, s)$\;
    }
  }
  \caption{Mapping experiment outline}
  \label{algo:mapoutline}
\end{algorithm}

\noindent We explain the functions referenced in Algorithm~\ref{algo:mapoutline} below:

\emph{generateIntialMapping} generates an initially mapped set for \emph{s} by removing a random number of randomly selected source code entities from the known mapping. It ensures that at least one source code entity is mapped to each module, so the initially mapped set will never be smaller than the number of modules in the intended architecture.

\emph{setParameters} sets the parameters of an attraction function \emph{af}, such as whether CDA should be used or not. The parameters can either be fixed or randomly selected from a range, e.g., random in the interval $[0.0, 1.0]$ or random Boolean. \change{Note that parameters are often unique for a specific attraction functions, e.g., only CountAttract uses $\omega$ and $\phi$ and CDA does not apply for CountAttract.}

\emph{initialize} initializes \emph{af} with an initially mapped set \emph{M}. The required initialization differs between the various attraction functions, e.g., creating the term documents for the initially mapped source code entities and the modules or training the Naive Bayes is trained.

\emph{findOrphans} returns the set of orphans for a system $s$ with an initially mapped set $M$. This is generally the set of source code entities except those initially mapped, but an attraction function might require additional filtering. For CountAttract, we filter the set of orphans as described in \cite{Christl2007AutomatedMethod} and Section~\ref{sec:background}. The other attraction functions do not use any filtering.

\emph{calculateAttractions} calculates the attraction of an orphan \emph{o} to all modules in the intended architecture using the attraction function \emph{af}. The attraction values are stored as part of \emph{o}.

\emph{candidateModule} decides if there is a single module \emph{m} that \emph{o} should be mapped to and, if so, returns that single module \emph{m}.

\emph{saveData} saves the experiment data, such as initial parameters for the attraction function, the size of the initial set, the number of failed and successful mappings, etc.

\begin{table}[pos=tb]
  \centering
  \caption{Subject systems and versions. $\mathbf{|E|}$, $\mathbf{|ED|}$, $\mathbf{|M|}$, and $\mathbf{|MD|}$ are the number of entities, modules, and respective dependencies between them. LoC are the Lines of Code in each subject system.}
  \label{tab:systems_overview}
  \setlength\tabcolsep{4.0pt}
  \begin{tabular}{@{}llrrrrrr@{}}
  \toprule
  \multicolumn{1}{c}{\textbf{Name}} & \multicolumn{1}{c}{\textbf{Version}} & \multicolumn{1}{c}{\textbf{LoC}}  & \multicolumn{1}{c}{$\mathbf{|E|}$} & \multicolumn{1}{c}{$\mathbf{|ED|}$} & \multicolumn{1}{c}{$\mathbf{|M|}$} & \multicolumn{1}{c}{$\mathbf{|MD|}$} \tabularnewline
  \midrule
  Ant       & r584500  & 36\,699 & 515   & 2\,476 & 16 & 86  \tabularnewline
  A.UML   	& r13713   & 62\,392 & 1\,485 & 8\,121 & 19 & 79  \tabularnewline
  C Img     & v1.0a2   & 17\,803 & 329   & 1\,485 & 21 & 21  \tabularnewline
  JabRef    & v3.7     & 59\,235 & 1\,015 & 5\,482 & 6  & 15  \tabularnewline
  Lucene    & r1075001 & 35\,812 & 515   & 2\,764 & 7  & 16  \tabularnewline
  T.Mates   & v5.11    & 54\,904 & 779   & 5\,841 & 15 & 28  \tabularnewline
  S H 3D    & r002382  & 34\,964 & 167   & 1\,175 & 9  & 29  \tabularnewline
  ProM      & v6.9     & 9\,947  & 261   & 747   & 4  & 5   \tabularnewline
  \bottomrule
  \end{tabular}
\end{table}

\subsection{Subject Systems}

We \change{study} eight open-source systems implemented in Java. Ant\footnote{\url{https://ant.apache.org}} is an API and command-line tool for process automation. Commons Imaging\footnote{\url{https://commons.apache.org/proper/commons-imaging/}} (C Img) is a library for reading and writing different image formats. ArgoUML\footnote{\url{http://argouml.tigris.org}} (A.UML) is a desktop application for UML modeling. JabRef\footnote{\url{https://jabref.org}} is a desktop application for managing bibliographical references. Lucene\footnote{\url{https://lucene.apache.org}} is an indexing and search library. ProM\footnote{\url{http://www.promtools.org/}} is an extensible framework that supports a variety of process mining techniques. Teammates\footnote{\url{https://teammatesv4.appspot.com}} (T.Mates) is a web application for handling student peer reviews and feedback. Sweet Home 3D\footnote{\url{http://www.sweethome3d.com}} (SH 3D) is an interior design application.

Table~\ref{tab:systems_overview} \change{presents the} version we evaluated and the sizes of the implementations and architectures, i.e., the number of entities, modules, and dependencies of each. For each system, there exists a documented architecture and a mapping from the implementation to this architecture. JabRef, Teammates, and ProM have been the studied at the Software Architecture Erosion and Architectural Consistency Workshop (SAEroCon) 2016, 2017, and 2019 respectively, where a system expert provided the architecture and the mapping. The architecture documentation and mappings are available in the SAEroCon repository\footnote{\url{https://github.com/sebastianherold/SAEroConRepo}}. Ant, ArgoUML, and Lucene were studied by \citet{brunet2012evolutionary} and \citet{LenhardExploringSourceCodeMetricsIndicatingArchInconsistency}, and the architectures and mappings were extracted from the replication package of~\citet{brunet2012evolutionary}. The mapping of Commons Imaging was provided in December 2020 by one of the current contributors as a test case for our tool, s4rdm3x.

\subsection{Data Analysis}

\change{
We measure the performance of each attraction function by precision and recall. For comparisons and visualizations, the precision and recall are combined to the harmonic mean F1 score. In addition to reporting the F1 score, we also report the overall precision and recall visually for each attraction function. This gives an overall view of the performance of each function regardless of the system.
}

The performance of CountAttract depends on the values of the $\omega$ and $\phi$ parameters. A system expert should ideally set them, but we use random values since we do not have access to \change{expert knowledge}. To make our experiment more realistic, we only take the best 50\% mappings in each 5\% segment of the initial set size. This ensures that we have roughly the same amount of data as for the other mapping functions and that we have data over the whole initial set size. We find that this results in a considerable improvement to the F1 scores of the CountAttract function, which improves our comparison's fairness.


We perform statistical significance testing to determine whether there is a significant difference in the F1 score and investigate the effect size. This is done per system and attraction function. We test for statistical significance at $p < 0.05$ using the Wilcoxon rank-sum test, as it is unlikely that the data is normally distributed \change{and the test is robust to outliers. As we do the statistical testing for the full range of initial set sizes, we do not want exceptionally good or bad data points to skew the results}. We report the effect size using $r$ and the median difference. As testing is done for the full interval of initial set sizes, it is interesting to show if there are specific intervals that are different visually. An attraction function may perform very well in a specific interval, giving an overall significant difference, but actually perform badly or equal to the other attraction functions in some other interval. Therefore, visualizations that show the median and the distribution of the F1 scores for the entire interval of the initial mapping sizes will complement the significance testing. 

First, we perform tests to assess the impact of using CDA in NBAttract, IRAttract, and LSIAttract. Then we evaluate the F1 scores of all attraction functions using the best data sets, with or without CDA, of NBAttract, IRAttract, and LSIAttract, respectively, and compare with CountAttract.

\section{Implementation of the Attraction Functions}\label{sec:implaf}

We implement the attraction functions in Java as part of our open-source tool suite for architectural analysis\footnote{Source code available at \url{https://github.com/tobias-dv-lnu/s4rdm3x}}~\cite{Olsson2021}.  
The implementation of CountAttract is based on the description in~\cite{Christl2007AutomatedMethod}, and the implementations of IRAttract and LSIAttract are based on the descriptions in~\cite{Bittencourt2010ImprovingTechniques}. Since the implementations are based on the textual descriptions and not source code, we cannot be certain that our implementations are correct, but \change{we find similar results provided in the publications to validate} the algorithms. We also validated our Java implementations against further independent implementations in Python that relied on the same descriptions. Results from our tools have been peer-reviewed~\cite{OlssonTowardsImprovedMapping, Olsson2021, OlssonNaiveBayes} and our implementations are available for inspection.

\subsection{Attraction Calculation\label{sec:afcalc}}

Our implementation of \emph{CountAttract} uses all available dependencies in the compiled source code. These are the following:

\begin{enumerate}[]
  \item \textit{Extends}, a type is inherited.
  \item \textit{Implements}, an interface is realized.
  \item \textit{Field}, a type is a member variable.
  \item \textit{Argument}, a type is an argument in an operation.
  \item \textit{Returns}, a type is used as a return value.
  \item \textit{LocalVar}, a type is used as a local variable.
  \item \textit{MethodCall}, an operation is called on a type.
  \item \textit{ConstructorCall}, a type is created.
  \item \textit{OwnFieldUse}, a type is used by a field in this type.
  \item \textit{FieldUse}, a type is used by a field in some other type.
  \item \textit{Throws}, a type is used in the form of an exception.
\end{enumerate}

We apply a per system unique weighting to these dependencies to further increase the performance of CountAttract. The weights were decided by optimization using a genetic algorithm approach. The results from the optimization were verified to perform better than setting all weights to $1.0$. \change{The possible values for the weights are limited to \{0, 0.25, 0.5, 0.75, 1\}. We use these values in the experiments to reflect that a human expert user would set them on a more coarse-grained scale than the actual, more precise optima.} Table~\ref{tab:dep_weights} shows the actual optimized weights; each is then rounded to one of the five values before they are used in the experiments. The genetic algorithm and its parameters are available in our replication package. 

The attraction values are evaluated according to the procedure described in~\cite{Christl2007AutomatedMethod}; if one and only one attraction value is higher than one standard deviation above the mean, or if one and only one attraction value is above the mean, the orphan is automatically mapped to the corresponding module. Variables $\omega$ and $\phi$ are assigned random values in the interval $[0.0, 1.0]$ with a uniform distribution.


\begin{table*}[]
\caption{Optimized weights per dependency and system.}
\label{tab:dep_weights}
\begin{tabular}{@{}lcccccccc@{}}

\toprule
         \multicolumn{1}{c}{\textbf{Dependency}}
       & \multicolumn{1}{c}{\textbf{Ant}}
       & \multicolumn{1}{c}{\textbf{A.UML}} 
       & \multicolumn{1}{c}{\textbf{C Img}}
       & \multicolumn{1}{c}{\textbf{Lucene}} 
       & \multicolumn{1}{c}{\textbf{ProM}}
       & \multicolumn{1}{c}{\textbf{JabRef}}
       & \multicolumn{1}{c}{\textbf{SH 3D}}
       & \multicolumn{1}{c}{\textbf{T.Mates}} \\ 
\midrule
Extends          & 1.00  & 0.57  & 0.64   & 0.92   & 0.34 & 1.00      & 0.78         & 1.00      \\
Implements       & 0.83  & 0.79  & 1.00   & 1.00   & 0.59 & 0.88      & 0.00         & 0.07      \\
Field            & 0.00  & 0.24  & 0.07   & 0.00   & 0.35 & 0.00      & 0.00         & 0.11      \\
Argument         & 0.00  & 0.31  & 0.16   & 0.00   & 0.18 & 0.00      & 0.00         & 0.00      \\
Returns          & 0.01  & 0.81  & 0.15   & 0.28   & 0.34 & 0.19      & 0.80         & 0.00      \\
Local Var        & 0.00  & 0.16  & 0.09   & 0.58   & 0.00 & 0.00      & 0.52         & 0.00      \\
Method Call      & 0.00  & 0.00  & 0.05   & 0.00   & 0.04 & 0.00      & 0.00         & 0.00      \\
Constructor Call & 0.00  & 1.00  & 0.06   & 0.70   & 1.00 & 0.69      & 0.81         & 0.00      \\
Own Field Use    & 0.00  & 0.00  & 0.00   & 0.00   & 0.00 & 0.00      & 0.00         & 0.00      \\
Field Use        & 0.41  & 0.00  & 0.87   & 0.30   & 0.01 & 0.00      & 0.00         & 0.00      \\
Throws           & 0.00  & 0.45  & 0.00   & 0.91   & 0.91 & 0.62      & 0.27         & 0.00      \\ \bottomrule
\end{tabular}
\end{table*}

\subsection{Term Generation}

IRAttract, LSIAttract, and NBAttract all rely on text extracted from the source code entities. We use the same procedure to \emph{generate terms} for all three. Terms are generated from package names, filenames (correspond to the outer class name in Java), method names, identifier names (attribute, argument, variable, and operations), and string constants. Our tool analyzes the compiled source code. Consequently, we do not have access to semantic information available in, for example, comments. We have performed preliminary experiments where we extract such information from the source code. But in these experiments, the additional information from comments did not improve the performance of any of the attraction functions. 

``CamelCasing'' is a common and often recommended approach to identifier naming in Java. Hence, we split all strings on capital letters, i.e., from ``camelCase'', we get two additional terms: ``camel'' and ``Case''. Accordingly, we also split based on ``Kebab-case'' and ``Snake\_case''. The resulting words are stemmed, using the Porter stemmer~\cite{PorterStemmer}. Finally, they are filtered based on length; we only consider words that are three characters or longer. 

We also add words corresponding to dependencies according to the CDA approach. We use the same set of dependencies as we used for CountAttract (cf.\ Section~\ref{sec:afcalc}).

Each orphan forms a document with terms from the compiled source code, and each module forms a document with terms from all its mapped concrete entities and the module name as an additional term.



\emph{IRAttract} employs a vector space model, where each term represents a vector component. The vector components are constructed using the term frequency with term frequency normalization. \citeauthor{Bittencourt2010ImprovingTechniques} advises against the commonly used TF-IDF~\cite{Bittencourt2010ImprovingTechniques}, which our initial experiments also confirm. We instead normalize using the maximum term frequency in each document. The final attraction is calculated using the cosine similarity function between an orphan's term vector and a module's term vector. The resulting attractions are evaluated using the same approach as in CountAttract. The use of CDA (or not) is a random Boolean parameter with a uniform distribution at 0.5.

\emph{LSIAttract} uses a term-document matrix, where each document (column) corresponds to a module, and the matrix components are constructed using raw term frequencies. The term-document matrix is then reduced using singular value decomposition to a low-rank approximation. Our implementation uses the standard Weka matrix libraries\change{~\cite{Weka}}, specifically \textit{SingularValueDecomposition}, to compute the LSI matrices. We use the number of modules, i.e., documents, as rank. The orphan document vector is then transformed into this lower-rank space, and the cosine similarity function is applied as in IRAttract. This produces attraction values that are evaluated using the same approach as in CountAttract. As before, the use of CDA (or not) is a random Boolean parameter with a uniform distribution at 0.5.

\emph{NBAttract} is implemented using the \textit{NaiveBayesMultinominal} classifier available in Weka\change{~\cite{Weka}}. We use the module documents to train the classifier. Our initial experiments showed that term occurrence, i.e., if a term is present or not, worked better than term frequency, so we compute the set of all terms in the initially mapped source code entities. We then compute the occurrence for each module and train the classifier based on that. We use the classification probabilities as attraction values, i.e., the attraction for an orphan to a module is the probability that the orphan belongs to the class that the module represents. 

When using CDA, we need to create text terms as if the orphan was mapped to a certain module and then perform the classification. This means that several classifications are needed when using CDA, so there is no guarantee that the sum of the attraction values will be 1. This poses a problem when deciding whether an orphan can be automatically mapped or not. Our tests showed a very low recall when using the method from CountAttract, so we decided on a threshold of 0.9 instead; if there is a single maximum attraction value and that exceeds the threshold, we automatically map the orphan to the module. The use of CDA (or not) is again a random Boolean parameter with uniform distribution at 0.5.

\section{Results}

We ran the experiments and collected at least 50\,000 samples for each system and technique, i.e., at least 200\,000 samples per system. The exact number of samples is slightly higher as our experiment tool runs a \change{batch} of experiments in a separate threads. All threads are allowed to finish successfully after 50\,000 executions are reached. In total, we collected 1\,401\,697 data points. Filtering for the CountAttract function removed 174\,125 data points. Tables \ref{tab:cda_data} and \ref{tab:comparison_data} present the exact number of data points for each data set. The raw data and the R scripts to analyze it are available in our replication package\footnote{\url{https://github.com/tobias-dv-lnu/Replication}}.

\change{Below we present the results of the experiment to answer: 1.\ How does CDA affect the performance of text-based attraction functions in general and 2.\ How well does our Naive Bayes attraction function perform compared to the state-of-the-art?}

\subsection{The Effect of CDA}

\begin{figure}[pos=tb]
\centering
\includegraphics[width=0.95\columnwidth]{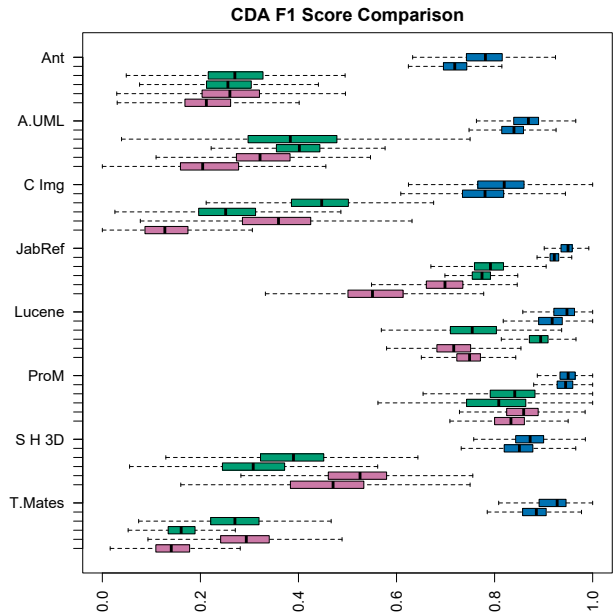}
\caption{Comparison of the NBAttract, IRAttract, and LSIAttract attraction functions with and without CDA for every subject system. The first two plots for each system are NBAttract (blue), the middle two are IRAttract (green), and the last two are LSIAttract (\change{pink}). \change{The first plot for every function and system, respectively, uses CDA and the second plot does not. We have omitted 49\,500 outliers out of a total of 1\,200\,600 data points to improve clarity.} \label{fig:f1_cda}}
\end{figure}

\begin{figure*}[width=13cm,align=\centering]
\centering
\includegraphics[width=1.94\columnwidth]{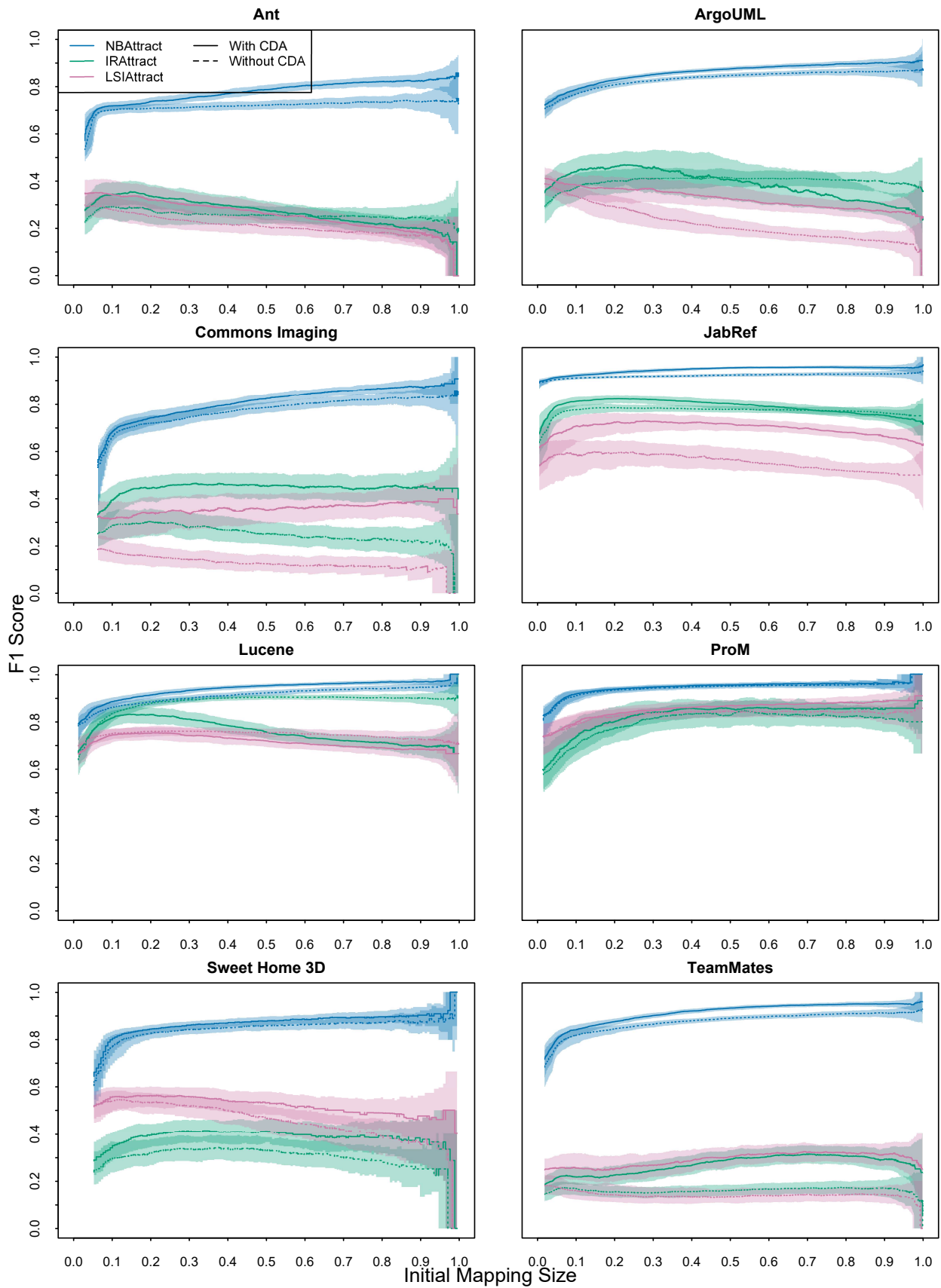}
\caption{The running median F1 score (vertical axis) with limits of the 75\textsuperscript{th} and 25\textsuperscript{th} percentiles for each system and attraction function with and without the CDA terms over the interval of initial mapping sizes (horizontal axis).\label{fig:f1_cda_plots}}
\end{figure*}

We compared the data for NBAttract, IRAttract, and LSIAttract concerning using the CDA terms or not. Table~\ref{tab:cda_data} reports the results of the statistical tests, where all comparisons are statistically significant ($p < 0.05$) in all cases except for one case. In ProM, there is no significant difference between using CDA or not for the NBAttract function. Figure~\ref{fig:f1_cda} shows box plots of the F1 scores for every system and attraction function. We find that using the CDA terms is beneficial in all cases except three: in IRAttract for Lucene and \change{A.UML}, and LSIAttract for Lucene. This is also confirmed in the statistical tests, see Table~\ref{tab:cda_data}. Figure~\ref{fig:f1_cda_plots} shows the running median mapping F1 scores, including limits of the 75\textsuperscript{th} and 25\textsuperscript{th} percentiles, over the interval of initial mapping sizes and systems. 


\begin{table}[pos=h!]
\centering
\caption{Results of statistical significance testing and effect size for comparing the F1 scores when using CDA terms or not. All results are statistically significant at $p<\mathrm{e}{-16}$. The $r$ statistic shows the standardized effect sizes and $\widetilde{\mathit{diff}}$ shows the median difference. A negative $\widetilde{\mathit{diff}}$ indicates that \change{using CDA terms is not beneficial} (marked with *). N1 shows the number of data points when using CDA and N2, respectively, when not using CDA.\label{tab:cda_data}}
\begin{tabular}{@{}llrrr@{}}
\toprule
\textbf{System} & \textbf{Statistic}          & \textbf{NB} & \textbf{IR} & \textbf{LSI} \\ \midrule    
Ant             & N1                          & 25\,061      & 25\,066      & 25\,109       \\             
                & N2                          & 24\,952      & 24\,947      & 24\,904       \\             
                & r                           & 0.54        & 0.09        & 0.27         \\             
                & Z                           & 121.15      & 19.03       & 60.54        \\             
                & $\widetilde{\mathit{diff}}$ & 0.06        & 0.01        & 0.05         \\             
\midrule
ArgoUML         & N1                          & 25\,085      & 24\,893      & 24\,972       \\             
                & N2                          & 24\,949      & 25\,141      & 25\,062       \\             
                & r                           & 0.35        & 0.05        & 0.54         \\             
                & Z                           & 77.47       & -11.10      & 121.80       \\             
                & $\widetilde{\mathit{diff}}$ & 0.03        & -0.02*      & 0.12         \\             
\midrule
Commons Imaging & N1                          & 24\,958      & 24\,969      & 25\,028       \\             
                & N2                          & 25\,056      & 25\,045      & 24\,986       \\             
                & r                           & 0.26        & 0.73        & 0.78         \\             
                & Z                           & 58.90       & 162.42      & 173.77       \\             
                & $\widetilde{\mathit{diff}}$ & 0.04        & 0.20        & 0.23         \\             
\midrule
JabRef          & N1                          & 24\,949      & 24\,828      & 24\,957       \\             
                & N2                          & 25\,068      & 25\,189      & 25\,060       \\             
                & r                           & 0.59        & 0.22        & 0.72         \\             
                & Z                           & 131.69      & 50.24       & 160.29       \\             
                & $\widetilde{\mathit{diff}}$ & 0.03        & 0.02        & 0.15         \\             
\midrule
Lucene          & N1                          & 25\,011      & 25\,045      & 24\,864       \\             
                & N2                          & 24\,996      & 24\,962      & 25\,143       \\             
                & r                           & 0.37        & 0.74        & 0.30         \\             
                & Z                           & 83.29       & -165.29     & -67.18       \\             
                & $\widetilde{\mathit{diff}}$ & 0.03        & -0.14*      & -0.03*       \\             
\midrule
ProM            & N1                          & 25\,045      & 24\,944      & 25\,030       \\             
                & N2                          & 24\,958      & 25\,059      & 24\,973       \\             
                & r                           & 0.10        & 0.17        & 0.26         \\             
                & Z                           & 22.50       & 38.70       & 58.16        \\             
                & $\widetilde{\mathit{diff}}$ & 0.01        & 0.03        & 0.03         \\             
\midrule
Sweet Home 3D   & N1                          & 24\,985      & 25\,116      & 24\,938       \\             
                & N2                          & 25\,029      & 24\,898      & 25\,076       \\             
                & r                           & 0.23        & 0.37        & 0.28         \\             
                & Z                           & 51.07       & 82.62       & 62.92        \\             
                & $\widetilde{\mathit{diff}}$ & 0.02        & 0.08        & 0.06         \\             
\midrule
TeamMates       & N1                          & 25\,055      & 24\,994      & 24\,823       \\             
                & N2                          & 25\,043      & 25\,104      & 25\,275       \\             
                & r                           & 0.43        & 0.69        & 0.76         \\             
                & Z                           & 97.03       & 154.65      & 169.20       \\             
                & $\widetilde{\mathit{diff}}$ & 0.04        & 0.11        & 0.15         \\ \bottomrule 
\end{tabular}
\end{table}

\subsection{Attraction Functions Comparison}

\begin{figure}[pos=t]
\centering
\includegraphics[width=0.95\columnwidth]{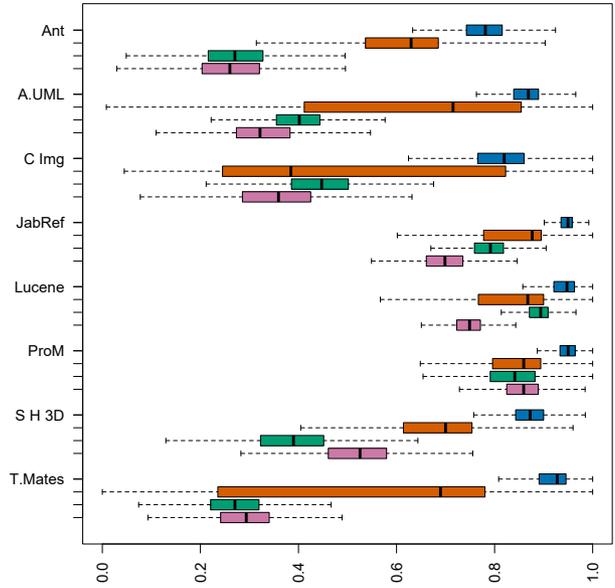}
\caption{Comparison of the NBAttract (blue), CountAttract (orange), IRAttract (green) and LSIAttract (pink) functions for each system. \change{We have omitted 42\,608 of 803\,047 data points for clarity.}\label{fig:f1_comparisons_boxplots}}
\end{figure}

\begin{figure}[pos=t]
\centering
\includegraphics[width=0.95\columnwidth]{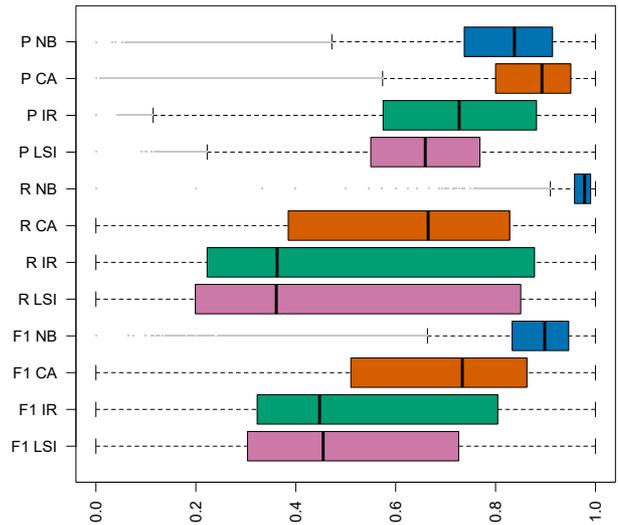}
\caption{\change{Comparison of the precisions, the recalls, and the F1 scores per attraction function: NBAttract (blue), CountAttract (orange), IRAttract \change{(green)}, and LSIAttract \change{(pink)} functions.} \label{fig:precision_recall_comparisons_boxplots}}
\end{figure}

\begin{figure*}[width=13cm,align=\centering]
\centering
\includegraphics[width=1.95\columnwidth]{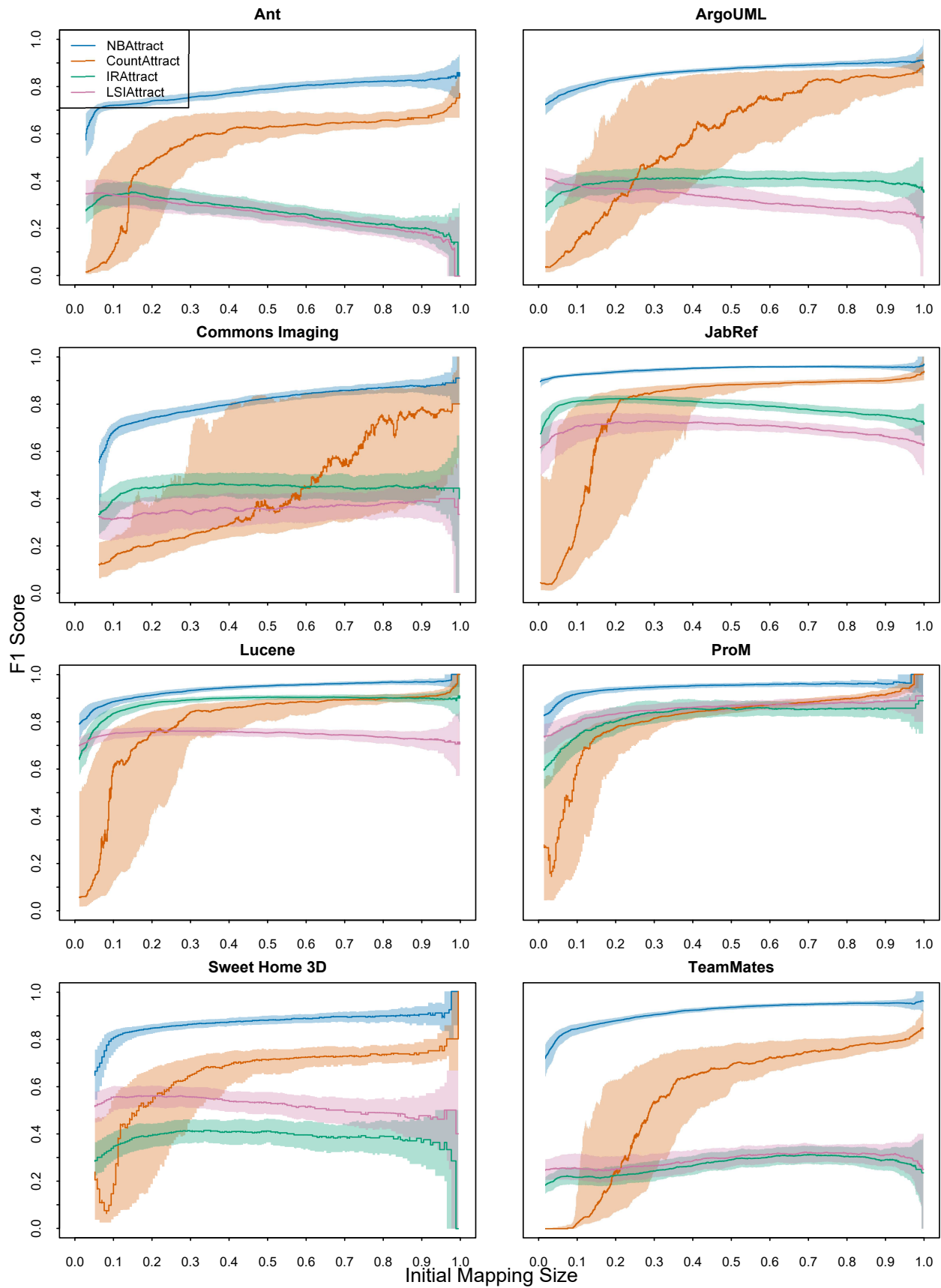}
\caption{The running median F1 score (vertical axis) with limits of the 75\textsuperscript{th} and 25\textsuperscript{th} percentiles for each system and attraction function over the interval of initial mapping sizes (horizontal axis).\label{fig:f1_comparison_plots}}
\end{figure*}

We compare the attraction functions: NBAttract, CountAttract, IRAttract, and LSIAttract. For NBAttract, IRAttract, and LSIAttract, we opt to use the best versions regarding CDA terms. Based on how effective CDA terms are, they are used in all cases except for IRAttract in the Lucene and ProM comparisons and LSIAttract for the Lucene comparison. For CountAttract, we apply the data point filtering previously described to simulate expert picks of the $\omega$ and $\phi$ parameters. Table~\ref{tab:comparison_data} shows the results of the statistical tests, where all comparisons are statistically significant (p < 0.05). Figure~\ref{fig:f1_comparisons_boxplots} shows boxplots of the distribution of F1 score across all subject systems and attraction functions. Figure~\ref{fig:f1_comparison_plots} shows the running median F1 score across set sizes, including the 75\textsuperscript{th} and 25\textsuperscript{th} percentiles. \change{Finally, Figure~\ref{fig:precision_recall_comparisons_boxplots} shows the overall precision, recall, and F1 score for each attraction function.}

\begin{table*}[]
\caption{Results of statistical significance testing and effect size for comparing all attraction functions. All results are statistically significant at $p<0.05$. The $r$ statistic shows the standardized effect sizes and $\widetilde{\mathit{diff}}$ shows the median difference. A negative $\widetilde{\mathit{diff}}$ indicates that the second function has a higher F1 score (marked with *). N1 shows the number of data points for the first function and N2 for the second function.\label{tab:comparison_data}}
\begin{tabular}{@{}llrrrrrr@{}}
\toprule
\textbf{System} & \textbf{Statistic}          & \textbf{NB vs. CA} & \textbf{NB vs. IR} & \textbf{NB vs. LSI} & \textbf{CA vs. IR} & \textbf{CA vs. LSI} & \textbf{IR vs. LSI} \\ \midrule    
Ant             & N1                          & 25\,061             & 25\,061             & 25\,061              & 25\,029             & 25\,029              & 25\,066              \\             
                & N2                          & 25\,029             & 25\,066             & 25\,109              & 25\,066             & 25\,109              & 25\,109              \\             
                & r                           & 0.47               & 0.86               & 0.86                & 0.66               & 0.66                & 0.10                \\             
                & Z                           & 106.30             & 192.22             & 192.58              & 146.76             & 148.44              & 22.96               \\             
                & $\widetilde{\mathit{diff}}$ & 0.09               & 0.40               & 0.42                & 0.32               & 0.33                & 0.02                \\             
\midrule
ArgoUML         & N1                          & 25\,085             & 25\,085             & 25\,085              & 25\,032             & 25\,032              & 25\,141              \\             
                & N2                          & 25\,032             & 25\,141             & 24\,972              & 25\,141             & 24\,972              & 24\,972              \\             
                & r                           & 0.28               & 0.86               & 0.86                & 0.47               & 0.53                & 0.41                \\             
                & Z                           & 63.35              & 192.90             & 193.02              & 105.63             & 119.59              & 91.60               \\             
                & $\widetilde{\mathit{diff}}$ & 0.09               & 0.42               & 0.49                & 0.33               & 0.40                & 0.07                \\             
\midrule
Commons Imaging           & N1                          & 24\,958             & 24\,958             & 24\,958              & 25\,059             & 25\,059              & 24\,969              \\             
                & N2                          & 25\,059             & 24\,969             & 25\,028              & 24\,969             & 25\,028              & 25\,028              \\             
                & r                           & 0.32               & 0.79               & 0.83                & 0.03               & 0.13                & 0.40                \\             
                & Z                           & 70.82              & 176.35             & 185.09              & -6.77              & 27.97               & 89.46               \\             
                & $\widetilde{\mathit{diff}}$ & 0.33               & 0.27               & 0.36                & -0.06*             & 0.02                & 0.08                \\             
 \midrule
JabRef          & N1                          & 24\,949             & 24\,949             & 24\,949              & 25\,019             & 25\,019              & 24\,828              \\             
                & N2                          & 25\,019             & 24\,828             & 24\,957              & 24\,828             & 24\,957              & 24\,957              \\             
                & r                           & 0.75               & 0.86               & 0.86                & 0.46               & 0.55                & 0.72                \\             
                & Z                           & 166.68             & 191.16             & 192.56              & 103.69             & 122.25              & 159.81              \\             
                & $\widetilde{\mathit{diff}}$ & 0.09               & 0.23               & 0.35                & 0.15               & 0.27                & 0.12                \\             
\midrule
Lucene          & N1                          & 25\,011             & 25\,011             & 25\,011              & 25\,058             & 25\,058              & 24\,962              \\             
                & N2                          & 25\,058             & 24\,962             & 25\,143              & 24\,962             & 25\,143              & 25\,143              \\             
                & r                           & 0.58               & 0.41               & 0.83                & 0.37               & 0.50                & 0.78                \\             
                & Z                           & 130.62             & 91.36              & 185.99              & -82.85             & 112.29              & 174.09              \\             
                & $\widetilde{\mathit{diff}}$ & 0.10               & 0.05               & 0.26                & -0.05*             & 0.16                & 0.22                \\             
\midrule
ProM            & N1                          & 24\,958             & 24\,958             & 24\,958              & 25\,109             & 25\,109              & 25\,059              \\             
                & N2                          & 25\,109             & 25\,059             & 25\,030              & 25\,059             & 25\,030              & 25\,030              \\             
                & r                           & 0.54               & 0.71               & 0.72                & 0.28               & 0.29                & 0.03                \\             
                & Z                           & 121.35             & 159.44             & 160.97              & 63.49              & 65.52               & -7.69               \\             
                & $\widetilde{\mathit{diff}}$ & 0.08               & 0.16               & 0.15                & 0.07               & 0.07                & -0.00*              \\             
\midrule             
Sweet Home 3D    & N1                          & 24\,985             & 24\,985             & 24\,985              & 25\,090             & 25\,090              & 25\,116              \\             
                & N2                          & 25\,090             & 25\,116             & 24\,938              & 25\,116             & 24\,938              & 24\,938              \\             
                & r                           & 0.65               & 0.85               & 0.84                & 0.65               & 0.56                & 0.55                \\             
                & Z                           & 145.23             & 190.48             & 186.73              & 146.65             & 125.84              & -123.44             \\             
                & $\widetilde{\mathit{diff}}$ & 0.15               & 0.49               & 0.37                & 0.34               & 0.22                & -0.12*              \\             
\midrule
TeamMates       & N1                          & 25\,276             & 25\,276             & 25\,276              & 25\,970             & 25\,970              & 24\,959              \\             
                & N2                          & 25\,970             & 24\,959             & 25\,294              & 24\,959             & 25\,294              & 25\,294              \\             
                & r                           & 0.70               & 0.86               & 0.86                & 0.49               & 0.48                & 0.14                \\             
                & Z                           & 159.06             & 193.61             & 194.13              & 111.18             & 108.92              & -32.33              \\             
                & $\widetilde{\mathit{diff}}$ & 0.18               & 0.62               & 0.60                & 0.44               & 0.42                & -0.02*              \\ \bottomrule 
\end{tabular}
\end{table*}

\section{Analysis}

Using CDA terms \change{seems beneficial in all but three cases: using the IRAttract function in Lucene and ArgoUML, and the LSIAttract function in Lucene}. A possible explanation is that there are often quite many CDA terms, which could become a dominating factor. IRAttract, especially, has a pretty simple approach to dealing with terms that dominate (i.e., a simple normalization). NBAttract inherently reduces the importance of terms that contribute less to the classification. The effect of using CDA varies among the systems, it is clearly positive in \change{thirteen} cases where the median CDA is above the 75\textsuperscript{th} percentile of its non-CDA counterpart (cf.\ Figure~\ref{fig:f1_cda}). In \change{eight} cases, the effect is positive, but below the 75\textsuperscript{th} percentile, \change{resulting in a positive effect in 21 of 24 cases.} Lucene and ProM systems stand out, where the \change{IRAttract and LSIAttract} functions produce either negative or below the 75\textsuperscript{th} percentile. \change{At the same time}, Lucene and ProM seem to be the ones that produce the highest and most consistent mapping results. This indicates that the entities have consistent terms according to the architecture. In such cases, finding the exact terms for each module would likely give superior results, while mixing various information sources would lower the results. The challenge in such cases is to be sure of the contributing factors. In \change{Lucene and ProM}, naming is a good candidate. It can also be argued that a human expert user could quickly manually map entities that are easy to map, i.e., via naming, and define the initial set. It would thus be interesting to run the experiments without the name terms, as this could show the contribution of \change{these}.

Overall, using CDA provides a positive benefit. \change{However, in the overall trends for IRAttract in Figure~\ref{fig:f1_cda_plots}, we find that the overall performance using CDA for Ant, ArgoUML, and JabRef decreases as the initial set grows. This is an interesting phenomenon that strengthens the hypothesis that CDA is not beneficial if large term frequencies cannot be handled. However, more experiments are needed to understand this effect fully.} 

Considering the simplicity of generating CDA terms and using them in text-based classification techniques, we recommend using this approach if the classification technique has some mechanism of balancing the term frequencies so that the terms do not dominate the classification.

Regarding the overall attraction function performance, we find that NBAttract clearly outperforms the other functions in all systems throughout different initial set sizes. The median NBAttract score is above the 75\textsuperscript{th} quartile scores of all other attraction functions for all systems. \change{Also, it clearly has less variation which indicates that it is less sensitive to variations in the initial set, see figures~\ref{fig:f1_comparisons_boxplots} and~\ref{fig:precision_recall_comparisons_boxplots}}. Table~\ref{tab:median_relation} provides a summary of the relationships of the median F1 scores for each subject system. CountAttract outperforms \change{IRAttract and LSIAttract in six of eight} cases, and \change{IRAttract outperforms LSIAttract in five of eight cases}. However, CountAttract seems to perform worse if a small initial set size is used, cf.\ Figure~\ref{fig:f1_comparison_plots}. This is consistent with the intended use of CountAttract in an evolutionary scenario where a majority of the entities are already mapped \cite{Christl2005EquippingClustering, Christl2007AutomatedMethod}. It should also be noted that CountAttract is highly optimized with system-specific weights for dependencies. \change{As discussed, we used a genetic algorithm approach to find optimal values for the weights and then rounded these to a five-value scale, as shown in Table~\ref{tab:dep_weights}. In reality, the weights would need to be decided by a human user, and it is unlikely that they would be optimal.}

\change{In our experiment, we used a fixed threshold of 0.9 as the limit for performing a mapping using NBAttract. This seems to favor recall over precision (cf.\ Figure~\ref{fig:precision_recall_comparisons_boxplots}). The threshold can be adjusted to, in theory, find the desired balance between precision and recall. CountAttract offers a similar mechanic using the $\omega$ parameter. However, it is arguably more intuitive for an end-user to adjust a probability threshold \change{compared to adjusting the $\omega$ parameter}. It would be interesting to investigate the effect of the mapping threshold for NBAttract in an empirical setting.}

\begin{table}[]
\centering
\caption{Relationship of the median F1 scores for all subject systems and attraction functions.\label{tab:median_relation}}
\begin{tabular}{@{}lc@{}}
\toprule
\textbf{System}       & \textbf{Relation Between Median F1 Scores} \\ 
\midrule
Ant	            &\makebox[20pt][c]{NB}\makebox[10pt][c]{\textgreater}\makebox[20pt][c]{CA}\makebox[10pt][c]{\textgreater}\makebox[20pt][c]{IR}\makebox[10pt][c]{\textgreater}\makebox[20pt][c]{LSI} \\
ArgoUML     	&\makebox[20pt][c]{NB}\makebox[10pt][c]{\textgreater}\makebox[20pt][c]{CA}\makebox[10pt][c]{\textgreater}\makebox[20pt][c]{IR}\makebox[10pt][c]{\textgreater}\makebox[20pt][c]{LSI} \\
Commons Imaging	&\makebox[20pt][c]{NB}\makebox[10pt][c]{\textgreater}\makebox[20pt][c]{IR}\makebox[10pt][c]{\textgreater}\makebox[20pt][c]{CA}\makebox[10pt][c]{\textgreater}\makebox[20pt][c]{LSI} \\
JabRef	        &\makebox[20pt][c]{NB}\makebox[10pt][c]{\textgreater}\makebox[20pt][c]{CA}\makebox[10pt][c]{\textgreater}\makebox[20pt][c]{IR}\makebox[10pt][c]{\textgreater}\makebox[20pt][c]{LSI} \\
Lucene	        &\makebox[20pt][c]{NB}\makebox[10pt][c]{\textgreater}\makebox[20pt][c]{IR}\makebox[10pt][c]{\textgreater}\makebox[20pt][c]{CA}\makebox[10pt][c]{\textgreater}\makebox[20pt][c]{LSI} \\
ProM	        &\makebox[20pt][c]{NB}\makebox[10pt][c]{\textgreater}\makebox[20pt][c]{CA}\makebox[10pt][c]{\textgreater}\makebox[20pt][c]{LSI}\makebox[10pt][c]{\textgreater}\makebox[20pt][c]{IR} \\
Sweet Home 3D	&\makebox[20pt][c]{NB}\makebox[10pt][c]{\textgreater}\makebox[20pt][c]{CA}\makebox[10pt][c]{\textgreater}\makebox[20pt][c]{LSI}\makebox[10pt][c]{\textgreater}\makebox[20pt][c]{IR} \\
TeamMates	    &\makebox[20pt][c]{NB}\makebox[10pt][c]{\textgreater}\makebox[20pt][c]{CA}\makebox[10pt][c]{\textgreater}\makebox[20pt][c]{LSI}\makebox[10pt][c]{\textgreater}\makebox[20pt][c]{IR} \\  \bottomrule
\end{tabular}
\end{table}

All attraction functions have greater variability of scores near the beginning and end of the initial set intervals. This indicates that the score is dependent on the composition of the initial mapping. This is not surprising when the initial mapping is small, i.e., functions require a representative initial mapping. Nevertheless, the variability seems to be even greater when the initial mapping is large, which we find surprising. This indicates that some set of entities are notoriously hard to map correctly, and if these are not included in the initial set, they have a large impact on the score, \change{e.g., generating a precision of zero}. This effect can also explain the declining median F1 score in large initial sets for CountAttract, IRAttract, and LSIAttract in Sweet Home 3D and TeamMates, cf.\ Figure~\ref{fig:f1_comparison_plots}.



None of the attraction functions has a perfect median score for automatic mapping, and using the F1 score as an assessment metric could be misleading depending on the purpose. For example, it could be better to get very few but always correct mappings giving, e.g., good advice to support human mapping for the rest. In general, this is a problem of software engineering tools, and an intuitive mechanism to balance precision and recall is beneficial for an end-user. In NBAttract, precision and recall can be balanced by adjusting the classification threshold. As this is a probability threshold, it should be intuitive for most end-users to understand and manipulate to achieve the desired effect.

\section{Threats to Validity}

\paragraph{External Validity} We have limited our study to different types of systems implemented in Java. While our findings should generalize to similar object-oriented languages, we do not know how well they are transferable to other contexts. \change{In particular, the structure of systems implemented in Java is tightly bound to file and folder names, which is often reflected in architectural modules in the subject systems. A system that has a radically different structure would likely be harder to map correctly.
Most of the systems in our study are of small to medium size; one cannot assume that the approach will scale to large or very large systems. For example, a large system could be more eroded and contain a more diverse set of architectural decisions that could influence the cohesiveness of modules.}

A random initial set is probably not a good representation of a real-world initial set. A system expert would start by mapping everything easy to map, e.g., whole packages, and possibly find a set of representative concrete entities. This could potentially leave only the hard or uncertain mappings to an automated mapping algorithm, which might struggle with these. So, it is unlikely that our F1 scores reflect a real-world scenario. We do not consider this a problem for our study, since our focus is to compare the different approaches to automated mapping, and their ability to deal with easier and harder cases is part of our evaluation given that we use random initial sets.

We also do not do any error correction or other human intervention during mapping. A human would likely impact the automatic mapping score provided some mechanism to assess the automatic mapping's correctness easily. Unfortunately, no such mechanism currently exists. The effect of humans guiding on the automatic mapping could, at least partially, be assessed by comparing the attraction functions using a larger initial set.

\paragraph{Construct Validity}\change{We are using the overall F1 score to assess the performance of the respective attraction functions. This way of calculating the F1 score can hide the fact that the performance for a minority class can be inferior. We have indications that some source code entities are very hard to classify correctly. A possible cause might be that they are part of tiny modules. This could mean that an attraction function that is more effective at handling such minority classes gets a lower overall F1 score compared to a function that is very good at majority classes. Our current analysis does not say anything about such problems.}

\change{We statistically compare F1 scores over a large span of different initial sets. This naturally produces potential outliers as, for example, small initial sets tend to perform worse overall. We choose to not remove outliers but instead use a significance test that is robust against the effect of outliers. We also complement the statistical tests with visual plots over the full span of initial set sizes.
}

There may also be a problem in the ground truth mappings as these are done by humans. In our experience, system experts will not always agree on every mapping. This subjectivity is also observed in architecture recovery studies by~\citet{Naim2017ReconstructingFramework}. It is possible that some, by the techniques wrongly mapped entities, could be a starting point for further discussion of the ground truth mappings with system experts.

Four of the systems (Commons Imaging, JabRef, ProM, and TeamMates) have mappings done by system experts. The others (Ant, ArgoUML, Lucene, and Sweet Home 3D) have mappings created by researchers studying the systems documentation and implementations~\citep{brunet2012evolutionary}. In the latter cases, there is a higher risk of errors. While not validated by architects or developers, \citet{brunet2012evolutionary} states that their derived architectures and mappings are consistent when they analyze developer code commits and mailing-list discussions. They also find that the intent of the architectural modules is stable during the period of their study. This leads us to think that a majority of the mappings are of high quality.

\change{Parts of the implementation are likely eroded, and this may create hard situations for any algorithm (and a human) to map. Erosion would typically create less cohesive source code entities until a source code entity becomes a mix of responsibilities. If the source code is severely eroded, automatic mapping could be impossible, and this is indeed what SACC is trying to prevent in the first place. In general, the consensus is that SACC techniques should be used early and continuously~\cite{Ali2017ArchitectureRequirements}. The performance of an automatic mapping is likely affected by the erosion in the source code, and we need to understand this effect further. Indeed one could argue that an entity that cannot be mapped automatically is problematic to some degree as it is not as well defined as other entities. \citet{tzerpos1997orphan} find that in 13 out of 46 misclassifications, developers see a need for refactoring of the source code. Similarly, \citet{olsson2016evaluation} find that 76 out of 101 misclassifications point to architectural problems in the source code. Preferably a mapping method would let a human user inspect these hard to map cases.} 

\paragraph{Internal Validity} As we discussed in Section~\ref{sec:implaf}, our implementations of the attraction functions are based on our best understanding of the original publications. However, the IRAttract and particularly the LSIAttract functions are not well described in \cite{Bittencourt2010ImprovingTechniques}, so we cannot be certain that our implementations are a correct representation of their work. LSI is not entirely suitable to use as an attraction function. Usually, LSI is used to find matching documents where there is an immense number of documents and terms. In the context of automatic mapping, the number of documents is quite small: the number of modules in the architecture is often less than 20. As the description of how to initialize and to use LSI is not sufficiently detailed in \cite{Bittencourt2010ImprovingTechniques}, our implementation might not be as intended. Alternative implementations could, for example, treat every entity in the initial set as a document. But, this raises the question of how to calculate the attractions to a module, i.e., should the best match be used, the median, or the average?

We do, however, consider the possible differences a minor issue, since our implementations are able to reproduce similar results to those provided in the original publication. Also, the differences in performance between IRAttract, LSIAttract, and NBAttract are quite large, so it seems unlikely that this relationship would be changed by differences in our implementation. 

CountAttract is sensitive to the input parameters $\omega$ and $\phi$. While our data filtering strategy tries to fairly simulate a more intelligent choice compared to random, it is not certain that this would reflect a system expert's choice of parameters. In fact, it would be likely that an expert would run and quickly inspect several mappings and then choose the best one. We also assume that the expert would select roughly the optimal weights, but it is improbable that all parameters will be optimal in a real-world scenario.

From a statistical perspective, it is not surprising that we get very low $p$-values: our data sets are quite large, and any non-random difference would likely be statistically significant. We also have the issue of using the same data set in multiple comparisons, and ideally, the $p$-value limit should be adjusted. Nevertheless, as our $p$-values are very low, this would not make any difference for the significance testing. In our case, the effect size matters the most. This is numerically available as median difference, $r$-value, and visually in boxplots and detailed plots with running median and running upper and lower percentiles.

\section{Related Work}


\citeauthor{tzerpos1997orphan} introduces the orphan adoption problem in a software evolution context~\cite{tzerpos1997orphan}. They are interested in two particular problems: how to map a new entity into a module (orphan adoption) and how to accommodate structural changes (orphan kidnapping), i.e., deciding when to remap an already mapped entity. They collectively name these problems the orphan adoption problem.

\citeauthor{tzerpos1997orphan} present a method for automatic orphan adoption and evaluate it in three case studies. They use the naming, structure, and style criteria. They use the naming and structure criteria for new orphans and the style criteria only for kidnapped orphans. Hard-coded patterns are used to decide whether an orphan follows the naming criteria and a developer classification of module styles limited to two styles. This means, they do not make any automatic style classification of new orphans. 

An interesting observation in one of their cases, that focused on an industrial system under development, is that their technique suggested 46 entities to change mapping compared to the developers' own assignment. The developers agreed on this new suggested mapping in 33 cases, and the remaining 13 original mappings were considered valid but not optimal. In these remaining cases, the developers expressed that restructuring the entities was needed to motivate their inclusion in the original modules.

\citeauthor{bibi2016orphan} evaluated three supervised learning techniques for automatic mapping \cite{bibi2016orphan}. They evaluate Naive Bayes, k-Nearest Neighbors, and Neural Nets as classifiers of orphans based on structural criteria from \citet{tzerpos1997orphan} and compare these with the original structural approach. For Naive Bayes, their approach uses the number of dependencies between a module and the orphan divided by the total number of entities in the module as the posterior probability of the orphan given the module. 



In their evaluation, Naive Bayes, k-Nearest Neighbors, and the original approach produced the same results. For Naive Bayes, we hypothesize that their approach to computing the posterior probability is too simplistic and fails to recognize patterns in the training data. It is reduced to counting actual dependencies and dividing by possible dependencies, limiting their approach.

In contrast, NBAttract incorporates multiple terms from several of the original criteria, e.g., style is learned from the actual patterns of dependencies expressed by the CDA terms, and the semantics are expressed by the terms available in, e.g., identifiers. As such, it can be argued that the style and semantics criteria are also incorporated.

As described earlier, the original HuGMe attraction functions try to solve the orphan adoption problem using the structural criteria, i.e., dependency information. However, HuGMe incorporates two major improvements compared to \cite{tzerpos1997orphan}. First, it makes the process iterative with human intervention to help with hard to classify cases. Second, it uses the structural information available between the modules in terms of convergent and divergent dependencies~\cite{Christl2005EquippingClustering, Christl2007AutomatedMethod}. 

\citeauthor{Christl2007AutomatedMethod} evaluate HuGMe using four subject systems, three in~\citep{Christl2007AutomatedMethod} and one additional case in~\citep{Christl2005EquippingClustering}. These are a code clone detection written in Ada95, an ANSI-C compiler written in C, a two-player client-server Tetris game written in Java, and finally, a system for graph visualization written in Java. Their evaluation focuses on the method as a whole for different sizes of the initial set and different values for the $\omega$ and $\phi$ parameters, for the two attraction functions CountAttract and MQAttract. They find that the CountAttract method of calculating the attraction values seems better from all points of view, but especially regarding keeping the number of modules suggested in manual mapping as few as possible. This is a key aspect of a semi-automatic technique: supporting the human user as much as possible.


CountAttract relies on the idea that coupling between the modules should be minimized and internal cohesion within a module maximized, i.e., it assumes a low coupling and high cohesion design. While this is, in general, a good practice, it may not always be the goal of the developers. \citeauthor{Candela2016UsingRemodularization}~\citep{Candela2016UsingRemodularization} found that some architectural modules are likely created with  design goals at an architectural level other than high cohesion and low coupling. They investigated 100 open-source projects from the Maven repository and conclude that a high percentage of the studied systems shows low values for structural (call graph-based) and conceptual (text analysis-based) package cohesion and coupling. Developers also confirmed that other properties guide the creation of modules and that multiple goals must be considered. Additionally, they found instances where an exact rationale for a module was present, despite it violating the high cohesion and low coupling principle. 

\citet{Bittencourt2010ImprovingTechniques} present attraction functions IRAttract and LSIAttract based on information retrieval as previously described. The automatic mapping performance of the attraction functions are evaluated both separately and in combination with CountAttract and MQAttract~\citep{Christl2005EquippingClustering,Christl2007AutomatedMethod}. The evaluation is performed for four subject systems using simulated evolutionary software changes on the scale of one orphan, 2--6 orphans, and 10--100 orphans. They find that a combination of LSIAttract and CountAttract has the best overall F1 score. They find semi-automatic mapping to be a feasible approach and that the overhead in an evolutionary context will be reduced compared to the manual work required. Though, faults in automatic mapping need special attention and tool support.

Graph clustering algorithms aimed at Automatic Architectural Recovery (AAR) and (AAR) re-modularization is strongly related to the problem of mapping implementation to higher-level constructs~\cite{Garcia2013ATechniques, pollet2007towardsSARTax}. The idea of AAR clustering seems to be very suitable for SACC purposes as well as incorporating heterogeneous information in the clustering algorithm analysis~\citep{ShternClusteringMethodologiesSE}. One important difference compared to Reflexion modeling is that AAR is essentially limited to the actual implementation and its possible flaws. Using Reflexion modeling, the architect can express the architecture as needed, i.e. the intended architecture, and Reflexion modeling can support multiple views.

\citet{misra2012ClusteringUnifyingSyntaxSemantic} propose an approach that unifies syntactical and semantic features to recover a component architecture. The approach uses a combination of textual features, such as code comments, identifiers, etc., and syntactic features, such as inheritance and dependency information. The features are combined using automatically generated weights and then used to estimate the distance between the code-elements. These distances could be used to generate attraction values in HuGMe. They evaluate the approach on five subject systems and find that using all features performs best, if there is no approach to select the appropriate feature type explicitly.

\citet{Naim2017ReconstructingFramework} present an approach to recover a system's architecture by combining contextual and structural information with incremental developer feedback. The approach is based on a novel adaptation of spectral clustering that can cluster heterogeneous graphs, e.g., one graph based on lexical analysis and the other based on the dependencies. In this approach, developer feedback is needed to tailor the initial clustering results to clarify each cluster's semantic terms. Their evaluation with four professional developers finds that this feedback is necessary to achieve  acceptable recoveries of the architectures.
\section{Conclusions}
\change {We present a new attraction function for semi-automatic clustering of source code to architectural modules, NBAttract. It leverages multinomial Naive Bayes classification and can use the broader set of information available in the source code and the architecture model. NBAttract does not require the user to decide parameters for different values, such as weights for dependencies. We also introduce CDA to represent dependencies as text, which allows NBAttract to combine syntactical and semantic information.} 

\change {We evaluate how CDA affects the performance of text-based attraction functions in general and the performance of our Naive Bayes attraction function compared to the state-of-the-art.}

Our experiments show that using a \change{machine learning} approach to semi-automatic mapping is clearly advantageous in all subject systems. NBAttract has a higher median F1 score and a lower variance than CountAttract, IRAttract, and LSIAttract throughout the interval of initial mapping set sizes for all subject systems. This, coupled with readily available implementations and the low need of parameterization, only setting of classification threshold, makes it a prime candidate for tools that implement SACC using Reflexion modeling. It also provides a useful baseline for research into more advanced techniques.

We evaluated the usefulness of our CDA approach of incorporating dependency information in attraction functions based on textual information and found an overall positive contribution. The implementation of CDA is straightforward and conceptually easy, so we recommend adding it to text-based classification techniques in most cases. However, care should be taken so that the CDA terms do not become too dominant, as they do in two cases when using IRAttract and one case in LSIAttract. NBAttract shows an ability to adapt the value of the CDA terms if they are not useful for the actual mapping. Overall, IRAttract lacks this ability without manual adaptation by the user, e.g., by introducing additional weights. Therefore, we do not generally recommend CDA when using IRAttract.

CountAttract has several parameters that need to be set by a system expert. The different types of dependencies between entities need to be weighed. Our experiments show that these weights have a significant effect on the performance. This is, at least in part, due to CountAttract assuming a design based high cohesion. \change{The weights need to reflect if cohesion is intended only for some types of dependencies or only to some degree in a system. We initially experimented with binary weights, i.e., turning off some types of dependencies.} While this resulted in improved performance compared to no weights, the optimized weights perform significantly better. We consider it difficult to set $\omega$ and $\phi$ to good values, even for a system expert, and close to impossible to set good weights for all dependency types. While our experiments show that an optimization algorithm can find good weights, they also show that NBAttract, which does not need this step, outperforms the other approaches with ``optimal'' weights.

While Naive Bayes offers a significant improvement over previous attraction functions, several problems remain. The automatic mapping is lacking in some subject systems, and there is a general need for further improvements of precision. There are plenty of machine learning techniques for text classification~\cite{TextClassAlgoSurvey} to evaluate in the context of the orphan adoption problem. It would be interesting to see if there are more suitable alternatives than Naive Bayes and if the positive effect of CDA can be found. While our initial studies suggest that additional text from the source code, such as comments, does not improve the classification, further investigation is needed to confirm this. 

\change{In general, there is a need to experiment using a more diverse set of systems of larger size and preferably in other programming languages. It would be particularly interesting to test on large systems where the organization of the source code does not clearly reflect the modular architecture.}


Our experiments demonstrate that some entities are never mapped correctly unless they are part of the initial set. This suggests that some entities are harder to map automatically than others, or even impossible. We need to study these entities further, so we can, e.g., learn to identify them and ensure they are mapped by a human. We are particularly interested in investigating whether hard cases are edge cases or actually incorrectly mapped by human experts, as found in~\citet{tzerpos1997orphan}.

\change{
\section*{Acknowledgement}
We sincerely thank the guest editors and the three anonymous reviewers for their generous time in providing detailed comments and suggestions that helped us to improve this article. We also thank the Centre for Data Intensive Sciences and Applications at Linnaeus University for supporting the research.
}

\bibliographystyle{cas-model2-names}

\bibliography{main}

\end{document}